\documentclass{article}

\usepackage[nonatbib,preprint]{neurips_2020}

\usepackage[utf8]{inputenc} 
\usepackage[T1]{fontenc}    
\usepackage{hyperref}       
\usepackage{url}            
\usepackage{booktabs}       
\usepackage{amsfonts}       
\usepackage{nicefrac}       
\usepackage{microtype}      
\usepackage{lipsum}
\usepackage{graphicx}
\usepackage{color}
\usepackage{url}
\usepackage{amsmath}
\usepackage{amssymb}
\usepackage{caption}
\usepackage{mathtools}
\usepackage{booktabs}
\usepackage[utf8]{inputenc}
\usepackage{relsize}

\DeclareUnicodeCharacter{200E}{}
\pdfoutput=1

\title{Adversarial learning for product recommendation}

\author{
	Joel R. Bock \\
    La Mesa, CA 91942 \\
	\texttt{jbock@ucsd.edu} \\
	\And
	Akhilesh Maewal \\
	San Diego, CA 92130\\
	\texttt{amaewal@gmail.com} \\
}

\begin{document}
	\maketitle	
	\begin{abstract}
Product recommendation can be considered as a problem in data fusion---estimation of the
joint distribution between individuals, their behaviors, and goods or services of interest. \color{black} This work proposes a conditional, coupled generative adversarial network (\emph{{RecommenderGAN}}) 
  that learns to produce samples from a joint distribution between \emph{{(view, buy)}} 
   behaviors found in extremely sparse implicit feedback training data. 
User interaction is represented by two  matrices having binary-valued elements. In each matrix, nonzero values indicate whether a user viewed or bought a specific item in a given product category, respectively. By encoding actions in this manner, the model is able to represent entire, large scale product catalogs.
Conversion rate statistics 
computed on trained GAN output samples ranged from $1.323\%$ to $1.763\%$. These statistics are found to be significant in comparison to null hypothesis testing results. The results are shown comparable to published conversion rates aggregated across many industries and product types.
Our results are preliminary, however they suggest that the recommendations produced by the model may provide utility for consumers and digital retailers.
\end{abstract}


\section{Introduction}

Product recommendation can be considered as a problem in data fusion---that is, estimation of the
joint distribution between individuals, their behaviors, and~goods or services of relevance or interest~\cite{Gilula2006}. This distribution is used to create a list of recommended items to present to a consumer.
\color{black}

\textls[-15]{\textbf{{Business impact of recommendation.}} 
 Online retailing revenue continues to expand each year. The~largest online provider of goods and services (Amazon) reported 2019 gross revenue of \${280.5B}
 , an~increase of 20.4\% over the { previous year }
 ({\url{https://www.marketwatch.com/investing/stock/amzn/financials}}). Most sizable e-commerce companies use some type of recommendation algorithm to suggest additional items to their customers. The~Long Tail proposition asserts that by making consumers aware of rarely noticed products via recommendation, demand for these obscure items would increase, shifting the distribution of demand away from popular items, and~potentially creating a larger market overall~\cite{Anderson2006}. The~goal of personalized recommendation is to produce marginal profit from each customer. These~incremental sales are certainly non-trivial, accounting for approximately 35\% additional revenue for Amazon, and~75\% for Netflix by some estimates~\cite{MacKenzie2013}. Operating efficiencies within a digital enterprise can also be significantly improved. Netflix saves \${1B} per year in cost due to churn by employing personalization and recommendation~\cite{Gomez-Uribe2015}. 
}

\textbf{{Recommender systems.}} Recommendation algorithms act as  filters to distill very large amounts of data down to a select group of products personalized to match a user's preferences.  Filtering and ranking the recommendations is extremely important; marketing studies have suggested that too many choices can decrease consumer satisfaction
and suppress sales~\cite{Iyengar2001}. 
These algorithms can be categorized into a few basic strategies---(1) item- or content-based (return lists of popular items with similar attributes); (2) collaborative (recommend items based on preferences or behaviors of similar users), or~(3) some hybrid combination of the first~two.

\textls[-15]{
\textbf{{Related work: Neural recommmenders.}}
\color{black}
Neural or deep learning-based recommendation systems are abundantly represented in the research literature (see reviews in~References \cite{Batmaz2018,Zhang2019}). Deep models have the capacity to incorporate greater volumes of data of mixed types, extract features and express \textit{{user-item-score}
} statistical relationships as compared to classical techniques based on linear matrix decomposition~\cite{Brand2003}.
Examples of deep algorithms application to recommendation tasks include multilayer perceptrons (MLPs) \cite{He2017};  autoencoders~\cite{Sedhain2015}; recurrent neural networks (RNNs)  
\cite{Hidasi2015}; graph~neural networks
~\cite{Ying2018}; and generative adversarial networks (GANs) \cite{Yoo2017}. These models aim to predict a user's preference for new or unseen items from mappings relating \textit{user-item} (\cite{He2017,Sedhain2015}),  \textit{item-feature} (\cite{Ying2018}) or \textit{item-item} sequences (\cite {Hidasi2015,Yoo2017}).
\color{black}}

\textbf{{Contribution of present research.}} In this work, we apply a deep conditional, 
coupled generative adversarial network
to a new domain of application-product recommendation in an online retail setting. In~the context of previous GAN research~\cite{Goodfellow2014}, and~specifically in terms of recommender systems, there~are several novel aspects of the model and approach advanced in this research. These~include:
\begin{itemize} 
	
	\item \textls[-15]{\textit{Mapping}: Direct modeling of the joint distribution between product \textit{views} and \textit{buys} for a user segment;}

	\item \textit{Data structure 
	\& semantics}: Inputs to the trained generative model are (1) user segment and (2)~noise vectors; the outputs are matrices of coupled (\textit{view, buy}) predictions;
	
	\item \textit{Coverage}: Complete, large-scale product catalogs are represented in each generated distribution;
	
	\item \textit{Data compression}: Application of a linear encoding algorithm to very high-dimensional data vectors, enabling computation and ultimate decoding to product space;
	
	\item \textit{Commercial focus} on transaction (versus rating) for recommended products by~design.
	
\end{itemize}

\color{black}

The \textit{RecommenderGAN} addresses many ongoing challenges found in recommender systems. Novelty and sparsity of recommendations is not an issue; samples representing the entire distribution of products in a high-cardinality catalog can be generated. Cold start issues are mitigated as the system learns to generate a joint $(view,buy)$ distribution for user segments with minimal identifying information, and~is designed to include additional  behavioral or demographic data if~available.

The following sections describe our experimental methods, details of the model construction and training, statistical metrics proposed for evaluation and results obtained with the adversarial recommender system. Concluding remarks identify areas for improvement of this approach to be addressed in future investigations.
\color{black}

\section{Methods}

\subsection{Background-Generative Adversarial~Networks}
Generative adversarial networks (GANs) are deep models that learn to generate samples representing a data distribution $p_{data}(\textbf{x})$  \cite{Goodfellow2014}. A~GAN consists of two functions: a generator $G$ that converts samples $\textbf{z}$ from a prior distribution $p_{\textbf{z}}(\textbf{z})$ into candidate examples $G(\textbf{z})$; and a discriminator $D$ that looks at real samples from $p_{data}(\textbf{x})$ and those synthesized by $G$, and~estimates the probability that a particular example is authentic, or~fake. $G$ is trained to fool $D$ with artificial samples that appear to be from $p_{data}(\textbf{x})$. The~functions $G$ and $D$ therefore have adversarial objectives which are described by the minimax function used to adapt their respective parameters:
\begin{equation} \label{eqn:gan}
\underset{G}{\operatorname{min}} \, \underset{D}{\operatorname{max}} \, V(G,D) = \
\mathbb{E}_{\textbf{x} \sim p_{data}(\textbf{x})} [  \mathrm{log}(D(\textbf{x}))  ] + \
\mathbb{E}_{\textbf{z} \sim p_{\textbf{z}}(\textbf{z})} [  \mathrm{log}(1-D(G(\textbf{z})) ]. 
\end{equation}

In the game defined in Equation (\ref{eqn:gan}), the~discriminator tries to make $D(G(\textbf{z}))$ approach 0; the~generator tries to make this quantity approach unity~\cite{Goodfellow2014}.

\textbf{{Conditional GANs.}}
Additional information about the input data can be used to condition the GAN model. To~learn different segments of the target distribution, an~auxiliary input signal $\textbf{y}$ is presented to both the generator and discriminator functions. The~objective function for the conditional GAN~\cite{Mirza2014,Gauthier2014} becomes
\begin{eqnarray}
\underset{G}{\operatorname{min}} \, \underset{D}{\operatorname{max}} \, V(G,D) &=& \mathbb{E}_{\textbf{x,y} \sim p_{data}(\textbf{x,y})} [  \mathrm{log}(D(\textbf{x,y}))  ] \quad +  \nonumber \\
& & \mathbb{E}_{\textbf{y} \sim p_{\textbf{y}}, \textbf{z} \sim p_{\textbf{z}}(\textbf{z})} [  \mathrm{log}(1-D(G(\textbf{z,y}),\textbf{y}) ], 
\end{eqnarray}
where the joint density to be learned by the model is $p_{data}(\textbf{x,y})$.

\textls[-15]{{{Coupled GANs}.} Further generalization of the GAN idea was developed to learn a joint data distribution $p_{data}(\textbf{x}_{1}, \textbf{x}_{2})$ over multiple domains within an input space~\cite{Liu2016}. The~coupled GAN   includes two paired GAN models [$f_{1}{:} (G{1},D_{1}), f_{2}{:} (G{2},D_{2}$)], each of which is trained only on marginal distributions $p_{data}(\textbf{x}_{1})$,  $p_{data}(\textbf{x}_{2})$ from the constituent domains of the joint distribution. The~coupling mechanism shares weights between the lower layers of $G_{1}$ and $G_{2}$, and~the upper layers of $D_{1}$ and $D_{2}$. respectively. The~architectural location of the shared weights in each case corresponds to the proximity of the greatest degree of abstraction in the data (for $G$, nearest to the input latent space $\textbf{z}$; for $D$, near~the encoded semantics of class membership). By~constraining weights in this manner, the~joint distribution $p(v,b)$ between \textit{view} and \textit{buy}
behaviors can be learned from training~data. }

\subsection{Model~Architecture}
The present model design combines elements of both conditional and coupled GANs as described above.  The~constituent networks of the coupled GAN recommender
were realized in software using the Tensorflow and Keras~\cite{Abadi2016,Chollet2015} deep learning libraries. The~GAN models were developed by adaptation of baseline models provided in an open source repository~\cite{Linder-Noren2018}.

\begin{figure}[h!]
	\centering
	\includegraphics[width=0.6\linewidth]{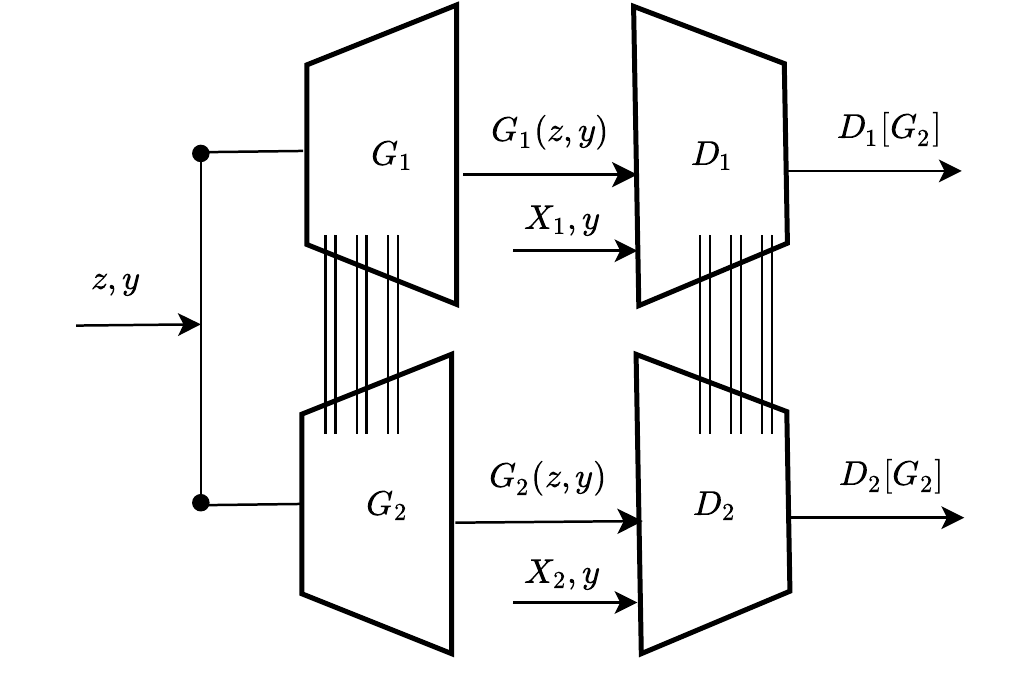}
	\caption{\textls[-15]{Coupled, conditional \textit{RecommenderGAN} architecture. All component networks $(D_{1}, D_{2},  G_{1}, G_{2})$ are active during training. Samples from the trained model are generated by presenting latent vector $z$ and user segment $y$ to the generators $G_{1},G_{2}$.}}
	\label{fig:gan}
\end{figure}

A schematic view of the
\textit{RecommenderGAN} model
used the current work is presented in
Figure~\ref{fig:gan}. The~generators (left) receive input latent vectors $z$ and user segment $y$, and~output matrices $G_{1}, G_{2}$. Discriminators (right) are trained alternatively with these artificial arrays and real samples $X_{1},X_{2}$, and~try to discern the difference. The~error in this decision is backpropagated to update weights in the~generators.

\textls[-15]{Architectural details of the network were settled upon after iterative experimentation and observation of results from different model design configurations and hyperparameter sets. Specifics of the layered configuration and dimensions of the final model appear in Table~\ref{tbl:architecture}.}

For the present dataset, more extensive (layer-wise) coupling of the weights within the generator networks proved necessary to obtain useful statistical results upon analysis. This is in contrast to the existing literature on coupled GANs, in~which the main application domain is computer vision~\cite{Liu2016}.

It was determined that the use of dropout layers~\cite{Srivastava2014} in $G_{1}, G_{2}$ improved convergence during training, but~such layers had negligible positive effect when included in the discriminator~sub-models.

The final model included a total of $326,614,406$ parameters, of~which $257,407,020$ were~trainable.

The optimization algorithm used for training the model was Adam (stochastic gradient descent), using default values for the {configurable parameters }({\url{https://keras.io/api/optimizers/adam/}).}

\color{black}

\begin{table}[h!]
	\caption{{Configuration details for} 
	discriminators (left) and generators (right). The~``?'' symbol indicates batch size of the associated tensor. Weights are shared between layers $D$:(9-14) and $G$:(6-13).}
	\begin{minipage}[t]{.4\linewidth}
		\vspace{0pt}
		\begin{tabular}{ccc}
			\toprule
			\textbf{ID} & \textbf{\textit{D} Layer}  & \textbf{Output Size}  \\
			\midrule
			1 & Input ($y$) & (?,1)  \\
			2 & Embedding & (?,1,500700) \\
			3 & Flatten & (?,500700) \\
			4 & Reshape & (?,1669,300,1) \\
			5 & Input ($X$) & (?,1669,300,1) \\
			6 & Multiply (4,5) & (?,1669,300,1) \\
			7 & AvgPooling & (?,834,15,1) \\
			8 & Flatten & (?,125100) \\
			9,10 & Dense, ReLU & (?,512) \\
			11,12 & Dense, ReLU & (?,256) \\
			13,14 & Dense, ReLU & (?,64) \\
			15 & Dense & (?,1) \\
			\bottomrule
		\end{tabular}
	\end{minipage}%
	\hfill 
	\begin{minipage}[t]{.4\linewidth}
		\vspace{0pt}
		\begin{tabular}{ccc}
			\toprule
			\textbf{ID} &\textbf{ \textit{G} Layer} & \textbf{Output Size}  \\
			\midrule
			1 & Input ($y$) & (?,1)  \\
			2 & Embedding & (?,1,100) \\
			3 & Flatten & (?,100) \\
			4 & Input ($z$) & (?,100) \\
			5 & Multiply (3,4) & (?,100) \\
			6,7 & Dense, ReLU & (?,128) \\
			8 & BatchNorm. & (?,128) \\
			9 & Dropout & (?,128) \\
			10,11 & Dense, ReLU & (?,256) \\
			12 & BatchNorm. & (?,256) \\
			13 & Dropout & (?,256) \\
			14 & Dense & (?,500700) \\
			15 & Reshape & (?,1669,300,1) \\
			\bottomrule
		\end{tabular}
	\end{minipage} 
	\label{tbl:architecture}
\end{table}

\color{black}

\subsection{Data~Preparation}
\color{black}

\textls[-15]{\textbf{{Electronic commerce dataset}}.
The adversarially trained recommender model was developed using a dataset collected from {an online retailer} 
 ({ \url{https://www.kaggle.com/retailrocket/ecommerce-dataset}}). The~dataset  describes website visitors and their behaviors (view, ``add'' or buy items available for purchase); the attributes of these items; and a graph describing the hierarchy of categories to which each item belongs. Customers have been anonymized, identifiable only by a unique session number; all items, properties and categories were similarly hashed for confidentiality~reasons.}

\textls[-15]{The dataset comprises  
2,756,101 behavioral events: (2,664,312 views, 69,332 ``add to carts'', 
22,457~purchases) observed on 1,407,580 distinct visitors. There are 417,053 unique items represented in the~data.}

\textls[-15]{Three different schemes for joint distribution learning can be constructed from the behaviors \textit{view}, \textit{addtocart}, and~\textit{buy} in the dataset---$(view,add)$, $(add,buy)$, and~$(view,buy)$. The~$(view,buy)$ scheme was studied here because it directly connects viewing a raw recommendation and a corresponding~purchase.}

\textbf{{User segmentation}}. In~Reference \cite{Lehmann2012}, models of user engagement with digital services were developed based on metrics covering three aspects of observable variables: \textit{popularity}, \textit{activity} and \textit{loyalty}. Each of these areas and metrics suggest means for grouping users in an implicit feedback~situation.

In the current study, users were segmented based on counts of the number of interactions made within a session (this is referred to as ``click depth'' in~Reference \cite{Lehmann2012}). Each visitor/session in the dataset were assigned to one of five behavioral segments according to click depth, which increases in proportion to bin number. Visitor counts by segment for the $(view,buy)$ scheme are summarized in Table~\ref{tbl:segments}. 

Note that the first segment is empty in this table. User segmentation was based on the entire sample; for the $(view,buy)$ group, no paired data fell into the lowest click {depth count bin.} 
({At least two actions are required to \textit{view} and \textit{buy}, and 86\% of the entire sample comprised two or fewer interactions, and~total buys less than 1\%}). The~total sample size considered for the four segments was 10,591~visitors.

\begin{table}[h!]
	\centering
	\caption{Visitor counts by segment for scheme $(view,buy)$.}
	\label{tbl:segments}
	\begin{tabular}{cccccc}
		\toprule
		\textit{\textbf{Segment}} & \textbf{0} & \textbf{1} & \textbf{2} & \textbf{3} & \textbf{4} \\ \midrule
		\textit{Count} &	n/a & 309 & 1182 & 1510 & 7590 \\
		\bottomrule
	\end{tabular}
\end{table}

\textbf{{Compressed representation}}. The~e-commerce dataset contained 
417,053 distinct items in 1669 product categories. Data matrices covering the full-dimensional $category \times item$ space ($\mathbb{R}^{1669 \times 417,053}$) are prohibitively large for practical computation. This is exacerbated by the number of trainable parameters in the model (>257,000,000).

To address this cardinality issue, a~compressed representation of the product data was created using an arithmetic coding algorithm~\cite{MacKay2003}. For~each visit, training data matrices were constructed having fixed  dimensions of $1169$ rows (one product category per row) and $300$ columns (encoded bit strings representing items in corresponding category). This was the maximum encoded column dimension; for row encodings of lesser length, the~bit string is prepended with zeros. The~decoded result is identical regardless of length of this prefix; this is important in order to subsequently identify specific items recommended by the~system. 

\color{black}

The encoded, sparse data matrices profiling visitor behavior for \textit{Views} ($\textbf{V}$) and \textit{Buys} ($\textbf{B}$) can be expressed symbolically as:
\begin{equation}
\mathbf{V}= \begin{bmatrix} 
v_{11} & v_{12} & \dots \\
\vdots & \ddots & \\
v_{r1} &        & a_{rc} 
\end{bmatrix}, \: \mathbf{B} = \begin{bmatrix} 
b_{11} & b_{12} & \dots \\
\vdots & \ddots & \\
b_{r1} &        & b_{rc}, 
\end{bmatrix}                         \label{eqn:matrices}           
\end{equation}
where elements $v_{m,n}, b_{m,n} \in \{0,1\}$ are indicator variables denoting  whether a visitor interacts with  category `$m$' and  item   `$n$', and~(r {=} 1669 $\times$ c {=} 300) is the encoded matrix~dimension. 

GAN training is carried out in the compressed data space, and~final recommendations are read out after decoding to the full column dimension of all items comprising the product~catalog.


\subsection{Evaluation~Metrics}\label{sec:eval-metrics}
\color{black}

Assessments of recommendation systems in academic research often include utility statistics of the returned results (such as novelty, precision, sensitivity), or~ overall system computational efficiency~\cite{McNee2006,Castells2011,Koren2008}. 
To estimate the business value derived from deployed systems, effectiveness measures may be direct (conversion rates, click-through rates), or~inferred (increased subscriptions, overall revenue lift, for~example) \cite{Jannach2019}. 

In the implicit feedback situation considered here, recommendations are created from sampling a joint distribution of $(view,buy)$ behaviors. Consider the potential paths for system evaluation as suggested in Figure~\ref{fig:evaluation-paths}. In~the left-hand column are the paired training data 
$(V_{x}, B_{x})$; on the right, the~generated recommendations $(V_{z}, B_{z})$.

\begin{figure}[h!]
		\begin{center}
		\includegraphics[width=0.5\textwidth]{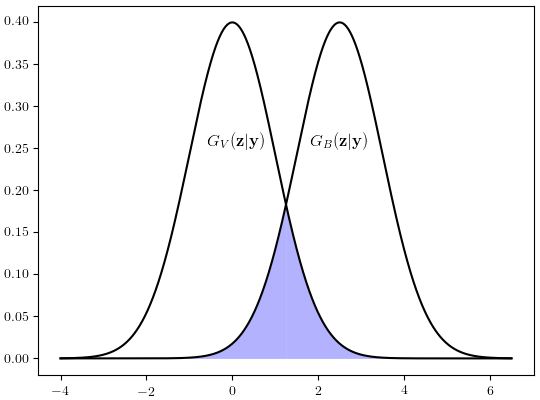}
			\end{center}
		\caption{Recommended items are drawn from intersection of outputs generated by the trained GAN, indicated in blue. $G_{V}(\mathbf{z}|\mathbf{y})$ and '$G_{B}(\mathbf{z}|\mathbf{y})$ are the  \textit{view} and \textit{buy} distributions, respectively.}
		\label{fig:overlap}
\end{figure}

Without knowledge of the relevance of recommendations by human ratings or via purchasing behavior, evaluation in this preliminary work is based on objective metrics of similarity between the generated $(view,buy)$ lists (path \#4). 
Items contained in the intersection of $(view,buy)$ generated recommendation sets are taken to signify the highest likelihood for completion of a transaction within the context of a given visitor session. This idea is illustrated in Figure~\ref{fig:overlap}.

\color{black}

Two metrics of evaluation are~proposed: 
\begin{enumerate}
	\item Specific  items contained within the overlapping category sets that are both viewed and ``bought''---a putative \textit{conversion rate};
	\item Coherence between categories in the paired $(view,buy)$ recommendations.
\end{enumerate}

Estimation of conversion rate is the most important statistic considered here; it is crucial for evaluation and optimization of recommender systems in terms of user utility, as~well as for  targeted advertising~\cite{Wen2019}.

Category overlap is prerequisite to demonstration of the feasibility of the current approach to product~recommendation.

\begin{figure}[h!]
	\begin{center}
		$\begin{array}[c]{ccc}
		V_{x}&\stackrel{1}{\Longleftrightarrow}&V_{z}\\
		\big\Updownarrow\scriptstyle{3}&&\big\Updownarrow\scriptstyle{4}\\
		B_{x}&\stackrel{2}{\Longleftrightarrow}&B_{z}
		\end{array}$
	\end{center}
	\caption{Evaluation routes for generative adversarial network (GAN)~recommender.}
	\label{fig:evaluation-paths}
\end{figure}
\unskip
\color{black}

\textbf{Product conversion rate.} Define the conversion rate as the number of items  recommended and bought, to~the count of all items recommended, conditioned on the set of overlapping product categories returned by the system:
\begin{equation}\label{eqn:cvr}
CVR=‎100 \times \frac{1}{N} \mathlarger{\mathlarger{‎\sum}}_{n=1}^{N}
\frac{\#(i_{v} \cap i_{b})}{\#i_{v} } \;
\bigg| \; \left\{ c_{v} \cap c_{b} \right\}_{y}, \quad (\%),
\end{equation}
where $(i_{v},i_{b})$ are items, $(c_{v},c_{b})$ are product categories, $N$ is the number of GAN realizations, $y$ denotes the user segment and ``$\#()$'' denotes  the cardinality of its argument. Note that in the current analysis, it~is assumed that all recommended items $i_{v}$ are viewed by a~visitor.

\textbf{Category similarity.} The average Jaccard similarity between recommended categories $(c_{v},c_{b})$ is given by
\begin{equation}‎\label{eqn:jaccard}
J_{c}=100 \times \frac{1}{N} \mathlarger{\mathlarger{‎\sum}}_{n=1}^{N}
\frac{\#(c_{v} \cap c_{b})}{\#(c_{v} \cup c_{b})},‎\quad (\%)‎.
\end{equation}

\textls[-15]{\textbf{Training distribution statistics.}
Summary statistics comparing the distributions $V_{x}, V_{z}$ (Figure~\ref{fig:evaluation-paths}, path~\#1) are observed to provide qualitative information about the effectiveness of target distribution~learning.
\color{black}}

\textls[-15]{\textbf{Null hypothesis tests.} A legitimate question to ask upon analyzing the current results is this: `\textit{`Are~the generator realizations samples of the target joint distribution, or~do they simply represent random~noise?}''.}

To address this question, the~analysis
includes statistics estimated from simulation trials ($n = 500$) in which randomly selected elements from the $(V,B)$ matrices are set equal to unit value, while~maintaining the average sparsity observed in the decoded GAN~predictions. 

The random trials are meant to test the null hypothesis that there is no correlation between paired $(view, buy)$ elements in the generator~output. 

The alternative hypothesis is that the recommendations contain relevant information that may provide utility to the system~user.

\color{black}


%
%
%
%
\color{black}

\subsection{Recommendation~Experiments} 

\textbf{Training.} The system was trained on the encoded data for 1100 epochs, in~randomly-selected batches of 16 examples~each.

Statistics of  training data for all networks comprising the model  were observed during the training~iterations.

\begin{figure}[h!]
	\centering
	\includegraphics[width=0.7\linewidth]{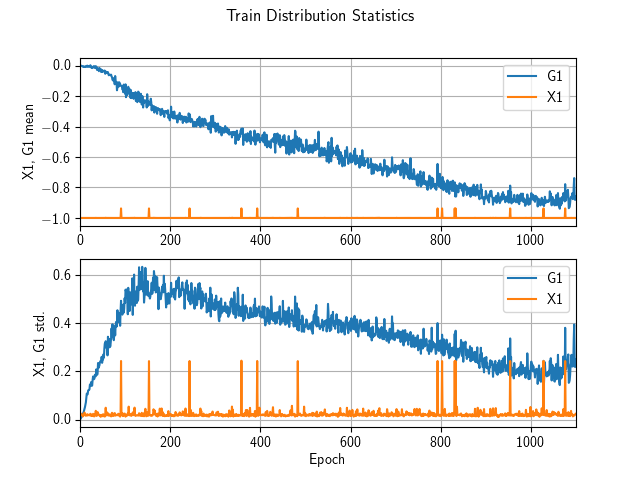}
	\caption{Batch statistics for $X_{1}$ and $G_{1}$ versus training epoch.
 Similar statistic values were observed for $X_{2}$, $G_{2}$. Batch size = 16.}
	\label{fig:training-curves}
\end{figure}

The $G_{1}, G_{2}$ statistics monotonically approached the true distribution those of the true  until around epoch nunber 1110,  at~which point the GAN output began to diverge. One explanation for this may be that the representational capacity of the networks on this abstract learning task may have become  exhausted~\cite{Bermeitinger2019}. Examples of this statistical evolution during training are shown in Figure~\ref{fig:training-curves}. Note that training data matrix values were scaled onto the range [$-$1,+1], where the value ``$-$1'' corresponds to a zero valued element in the sparse raw data arrays.
\color{black}

Label smoothing on the positive ground truth examples presented to the discriminators was used to regularize the learning procedure~\cite{Szegedy2015}. 

At training stoppage, the~observed discriminator accuracies where consistently in the $\approx 45{-}55\%$ range, indicating that these models were unable to differentiate between the real and fake distributions produced by the generators~\cite{Goodfellow2014}. 

\color{black}

\textbf{Testing.} Testing a machine learning model refers to evaluation of its output predictions obtained using out-of-sample (unseen) input data. This provides an estimate of the quality and generalization error of the model. Techniques such as  cross-validation are often used to assess generalization potential. In~contrast to many other learning algorithms, GANs 
do not have an objective function, rendering performance comparison of different models difficult~\cite{Salimans2016}.

For a concrete illustration, imagine that a GAN has been trained on a distribution of real images of human faces, and~generates synthetic face samples after sufficient training iterations~\cite{Karras2017}. 

The degree to which the sampled distribution has learned to approximate the target distribution can be estimated by qualitative scoring; the assessment is subjectively accomplished 
 by human observers,  who easily measure how ``facelike'' these artificial faces appear to the~eye. 

Alternatively, objective metrics based on training and generated image data can be applied in some cases. Example objective metrics are proposed in~Reference \cite{Salimans2016}.

In the present application, the~generated ``images'' are abstract representations of consumer activity, not concrete objects.  An~out-of-sample test in the conventional sense is not possible. The~metrics of generative model performance and null hypothesis tests as described in
Section~\ref{sec:eval-metrics} constitute the testing of the model developed in this~work.

\color{black}

\textbf{GAN predictions.} After training, the~model was stimulated with a noise vector and a user segment conditioning signal, producing a series of coupled $(view, buy)$ predictions $(G_{1}(z,y),G_{2}(z,y))$, as~depicted in Figure~\ref{fig:gan}. The~discriminators $D_{1},D_{2}$ serve only to guide training, and~are disabled for the inference~procedure.

A total of $2500$  generation realizations was produced for each user segment. The~recommendation matrices $V_{z}, V_{b}$ were decoded onto the full-dimensional $category \times item$ space ( $\mathbb{R}^{1669 \times 417,053}$ )
by the inverse arithmetic coding algorithm used in data~preparation. 

An $8\%$  {sub-sample } of these realizations was taken to compute key recommender evaluation metrics (Equations \ref{eqn:cvr} and \ref{eqn:jaccard}). 
This was done because the decoding algorithm involves compute-intensive iteration, and is not amenable for computation on GPUs at this writing. 
Numerical optimization of the arithmetic decoding should be addressed in future extensions of this work. 

The evaluation statistics were also calculated in null hypothesis tests, the~results of which were averaged to obtain an estimate of the distribution expected under this~hypothesis.

\section{Results And~Discussion}

\subsection{Main Statistical~Results.}
The main experimental results of this paper are summarized in Table~\ref{tab:results}. $CVR$ is the conversion rate (Equation (\ref{eqn:cvr})) and $J_{C}$ is the category similarity (Equation (\ref{eqn:jaccard})). Here, $y$ represents the user segment; $\#I,\#C$ are the predicted item and category counts, respectively. Row data in the table are averages over (1) 200 realizations from the GAN, and~(2) 500 random trials (columns marked ``$rn$''.)

The conversion rates ($CVR$) calculated from the GAN output range from $1.323$ to $1.763\%$. The~corresponding randomized values ($CVR^{rn}$) are at least 3 orders of magnitude smaller. Based on this result, the~null hypothesis of no relationship between paired $(view, buy)$ samples in the generator output is~rejected.

{{On this key statistical metric,} 
the~joint distribution of $(view,buy)$ behaviors sampled from the trained GAN model is characterized by non-trivial signal to noise. The~conversion rates observed from the sampled recommendations suggest utility of the system for consumers and digital retailers.}

\textls[-15]{The category similarity for the randomized cases is around $50\%$ for all user segments, while the GAN similarity is between $6.13$ and $8.19\%$. The~random procedure shows much greater similarity between recommended ($view$) and potential transactions--however given the negligible precision ($CVR^{rn}$), this~similarity metric has negligible practical~utility.}

\begin{table}[h!]
	\centering
	\caption{Average conversion rate and category similarity by~segment.}
	\begin{tabular}{ccccccc}
		\toprule
		\boldmath{$y$} & \boldmath{$\#I$}  & \boldmath{$\#C$} & \boldmath{$CVR$}   &\boldmath{$CVR^{rn}$} & \boldmath{$J_{c}$}    & \boldmath{$J_{c}^{rn}$} \\ \midrule
		1 & 1648 & 239 & \textbf{{1.763}} 
		& {0.0005}
		      & 8.19 & 50.66 \\
		2 & 2037 & 213 & \textbf{1.414} & 0.0004
		& 7.37 & 51.36 \\
		3 & 2522 & 190 & \textbf{1.323} & 0.0005     & 6.13 & 50.04 \\
		4 & 1419 & 222 & \textbf{1.644} &  0.0004      & 7.57 & 50.81 \\ \bottomrule
	\end{tabular}
	\label{tab:results}
\end{table}
\unskip

\subsection{Benchmark Comparison~Results.} 
To place the current results into commercial context, experimental conversion rates were compared against benchmarks observed by digital retailers across industries and product types. In~an online resource~\cite{Ogonowski2020}, conversion rates are estimated for 11 different industries and 9 product types. For~a concise comparison, average values of these conversion rates are shown along with the mean GAN conversion rate obtained here. The~data are summarized in Table~\ref{tab:cvr-benchmark}.

\textls[-15]{{It is seen from  Table~\ref{tab:cvr-benchmark} that the GAN conversion rates are slightly less than, but~on the order of, \linebreak aggregated industrial and product type values}. It is noted in~Reference \cite{Ogonowski2020} that precise definitions of ``conversion rate'' may vary and the one used in this research may be slightly different than industrial convention. Nevertheless, the~\textit{RecommenderGAN} produces intriguing rates of conversion given the paucity of information about user segments that was used to develop this~model.}

\begin{table}[h!]
	\centering
	\caption{Experimental conversion rate compared to selected industrial average benchmarks. \textit{GAN}: current results, segment-wise average; \textit{Industry}: average over 11 industrial markets; \textit{Product}: average of 9 product types~\cite{Ogonowski2020}.}
	\label{tab:cvr-benchmark}
	
	\begin{tabular}{ccc}
		\toprule
		\textbf{\textit{GAN}} & \textbf{\textit{Industry}} & \textbf{\textit{Product}} \\ \midrule
		\textbf{{1.536}} & 2.089 & 1.827 \\ \bottomrule
	\end{tabular}
\end{table}


\subsection{Discussion}
The results obtained from evaluation of the RecommenderGAN generated samples suggest that this approach may be useful for online recommendation systems.  This section discusses difficulties with direct comparison to other deep models, identifies areas for improvement of the model presented in this preliminary work, and~ other assorted~notes.

\subsubsection{Comparison with Other~Recommenders} 
\color{black}

A direct objective comparison of the present results with deep recommender systems found in the literature is problematic.

First, the~present scheme lacks ground truth data with which to evaluate novel recommendations produced by the \textit{RecommenderGAN}.
This fact motivated the design of the evaluation metrics discussed in Section~\ref{sec:eval-metrics}. GAN outputs are predictions that must ultimately be validated by human ratings of quality, or~via offline methods such as A/B testing
~\cite{Gilotte2018}. 

Secondly, the~GAN outputs a distribution of evenly-weighted recommended items, not numeric ratings that could be used to compare against other algorithms using rank-based statistical metrics.
\color{black}
Commonly used benchmarking datasets are based upon entities including \textit{user}, \textit{item}, \textit{rating} and some \textit{attributes}. In~order to compare against other deep models directly, it would be necessary to either (i)~refactor benchmark datasets from a rating scale for example, [1--5] for the MovieLens dataset~\cite{Harper2015}) onto the current binary-valued training data elements and re-interpret as a \textit{buy}/\textit{no buy} outcome, or~vice~versa (arbitrarily convert binary to a numeric scale).
Such modification would shift the meaning of the data and the model, and~it is not clear how to do this in a rigorously correct~manner.

%
%
%
%

A qualitative comparison of the \textit{RecommenderGAN} against selected deep recommendation models on the basis of their respective core algorithms,  inputs and outputs is presented in Table~\ref{tab:algo-comparison}.

\begin{table}[h!]
	\centering
	\caption{Comparison of present model and selected deep~recommenders.}
	\label{tab:algo-comparison}
	
	\begin{tabular}{cccc}
		\toprule
		\textbf{Algorithm} & \textbf{Reference} &
		\textbf{Recommender Input}  & 
		\textbf{Recommender Output} \\ 
		
		\midrule
		MLP+Matrix factorization & He~et~al. \cite{He2017}  & User, item vectors & Item ratings \\
		Autoencoder & Sedhain ~et~al. \cite{Sedhain2015} &  User, item vectors & Item ratings \\
		Recurrent neural network & Hidasi ~et~al. \cite{Hidasi2015} & Item sequence & Next item \\
		Graph neural network & Ying ~et~al. \cite{Ying2018} &  Item/feature graph & Top items \\
		Sequence GAN & Yoo ~et~al. \cite{Yoo2017} & Item sequence & Next item \\
		\textit{RecommenderGAN} & \textit{This work }  & \textit{Noise, user vectors} & \textit{View, buy  matrices}
		\\ \bottomrule
	\end{tabular}
\end{table}

\subsubsection{Drawbacks of Current~Method}
\color{black}

\textbf{Numerical efficiency.} A limitation of the approach to recommendation as presented here is the numerical efficiency of the decoding process. The~arithmetic coding algorithm used to decode the binary data matrices after training the model involves iteration and is not easily parallelizable. The~dimensionality of the full catalog of products is extremely high; decoding compute times are consequently large. This mandates offline processing before~deployment.

Future development should focus on numerical optimization of the decoding algorithm, to~perhaps include compilation to another language, such as C++.

\textls[-15]{\textbf{Ranking of recommendations.}  As discussed above, \color{black} there is no ranking of recommendation results in the current scheme, as~the GAN produces binary valued information upon decoding.
Inherent filtering is accomplished by limiting the presented results to those contained within the category intersection set $\left\{ c_{v} \cap c_{b} \right\}$ as seen in the operational definition of conversion rate (Equation (\ref{eqn:cvr})). This set is interpreted as representing the greatest likelihood for completing a transaction. On~average over user segments, $13\%$ of all categories are returned; of these, $0.46\%$ of all catalog items are~represented. 
}

\textls[-15]{\textbf{Conditioning signal.}
The current conditioning signal $y$ is simply based on user dwell time. The~information contained in this signal is relatively weak, as~indicated by the variation of statistics across segments in Table~\ref{tab:results}. It is reasonable to anticipate more stringent filtering, and~consequent precision and relevance of results, upon~the introduction of more robust demographic or behavioral data in the conditioning signal input to the model. This
would facilitate a more personalized recommendation experience.
The model  architecture considered here directly supports such segmentation, and~is an important topic to be explored in extensions to this research.
\color{black}
}

\subsubsection{General Discussion~Points}
\color{black}

\textbf{Selection bias and scalability.}
The estimation of conversion rates is difficult due to two related, \linebreak key issues: training sample selection bias and data sparsity~\cite{Wen2019}. Sample selection bias refers~to discrepancies in data distribution between model training and inference in conventional recommenders---that is,~training data often comprise only ``clicked'' samples, while inference is made on all impression samples. Selection bias is said to limit the accuracy of inference assuming the user proceeds through the sequence  $(impression \rightarrow click \rightarrow buy)$ \cite{Wen2019}. As~$clicked$ examples are a small fraction of $views$, a~highly imbalanced training set results, biased towards sparse positive examples~\cite{Hu2008}.

This issue is partially avoided in the present research, where the training data are constructed from all $viewed$ items, extracted from the user sequence  $(view \rightarrow addtocart \rightarrow buy)$. Model~recommendations are produced on items having semantic correspondence to the first and third actions in this sequence ({Note: Selection 
bias is not limited to implicit recommendation systems. Evaluating the performance of online recommenders with user feedback is biased towards the selectively recommend items~\cite{Longqi2018})}.

Intertwined with the data sparsity situation in implicit recommendation is the issue of scalability. In~collaborative filtering algorithms, the~consideration of all paired user-item data points as input to a model is infeasible; the numbers of those pairs can be exceedingly large, and~as noted, each user provides feedback on a very small fraction of the available items~\cite{Hu2008}.

The present system provides scalability to the full product catalog (417,053 items) by virtue of the arithmetic coding compression scheme, albeit at the cost of numerical performance upon~decoding.

\textls[-15]{\textbf{An open question.}
Has the true joint distribution been learned? Making inferences about the joint distribution of viewing and buying behavior to inform marketing decisions is the motivation behind this analysis. Investigators have previously shown that GANs may 
not adequately approximate the target distribution, as~the support of the generated 
distribution was low due to so-called \textit{mode collapse} \cite{Arora2017}, where the generator learns to mimic certain modes in the training data in order to trick the discriminator during~training.  }

It is debatable whether or not mode collapse is an issue in the present problem formulation, given the abstract formulation of the problem using binary indicator matrices to represent consumer behavior. This is an open question that may be addressed in further~research.

\section{Conclusions}

We have shown  that a coupled, conditional generative adversarial network can learn to generate samples from the joint distribution of online user behavior inferred from ($view,buy$) item pairs. These~samples can be used to make product recommendations for specific user segments.

Our results are preliminary, however they suggest that the recommendations produced by the model may provide utility for consumers and digital retailers.
\color{black}

Conversion rate statistics 
computed from the trained GAN output samples ranged from $1.323$ to $1.763\%$. These statistics were shown to be significant in comparison to null hypothesis testing results. This suggests that the GAN recommendations may provide utility for consumers and digital~retailers.

A comparison of GAN-predicted conversion rates against benchmarks from digital retailers representing many industries and product types showed the GAN conversion rates to approximate these aggregated commercial~rates.

The capacity of the system scales to the full item catalog dimension (417,053 items) by the use of an arithmetic coding compression~algorithm.

A number of original aspects of the approach to recommendation advanced in this research can be identified. These include---(i) Direct modeling of the joint distribution between product \textit{views} and \textit{buys} for a user segment; (ii) I/O: Inputs to the trained generative model are user segment and noise vectors; the outputs are matrices of coupled (\textit{view, buy}) predictions; (iii) Coverage of complete product catalogs can be represented in each generated distribution; (iv) Application of a linear encoding algorithm to very high-dimensional data vectors, enabling computation and ultimate decoding to product space;
(v) A focus on transaction (versus user rating) for recommended products by design.
\color{black}

 In future development of our model, a~means for objective comparison against competitive recommendation algorithms should be prioritzed. Such comparison should be carried out following a systematic methodology as  described in~Reference \cite{Dacrema2019}. This would require the articulation of a strategy for reconciling the (presently) binary-valued GAN outputs with numeric ratings in common usage with other recommendation algorithms. In~addition, work is needed to improve the numerical performance of the decoding algorithm used to invert to product space.
\color{black}


\bibliographystyle{unsrt}  
\bibliography{gan}

\end{document}